\documentclass[conference]{IEEEtran}
\IEEEoverridecommandlockouts

\usepackage{amsmath,amssymb,amsfonts}
\usepackage{graphicx}
\usepackage{booktabs}
\usepackage[hidelinks]{hyperref}
\graphicspath{{figures/}}

\newcommand{\bopt}{b_{\mathrm{opt}}}
\newcommand{\relMSE}{\mathrm{relMSE}}
\newcommand{\Wmax}{W_{\max}}
\newcommand{\bound}{\mathrm{bound}}

\begin{document}

\title{Closed-Form Optimal Weight Splitting for Matrix--Vector Multiply
on Noisy In-Memory Accelerators}

\author{\IEEEauthorblockN{Mehmet Aktukmak and David A. Roberts}
\IEEEauthorblockA{Micron Technology, Inc.}}

\maketitle

\begin{abstract}
Analog in-memory computing performs the matrix--vector multiplications of neural networks directly
inside the memory array, which saves time and energy. But each memory cell stores only a few levels
and is noisy, so one cell cannot hold a weight precisely. The standard fix is to \emph{split} each
weight across several cells, like digits of a number. The \emph{base} of this split decides the
trade-off: a large base gives fine resolution but amplifies the noise, and a base near one averages
the noise but wastes resolution. The best-known prior analysis considers only the noise, and so it
always picks a base of one. We propose a principled method that also accounts for the limited number
of levels per cell. It gives the optimal base in closed form, from a single measurement of the device
noise. On a device-physics simulator, the optimal split is 3.3~dB more accurate than the common
base-16 split, and it beats equal splitting in end-application accuracy.
\end{abstract}

\section{Introduction}\label{sec:intro}
The matrix--vector multiply (MVM) is the computational core of scientific
computing (the inner loop of Krylov solvers, the application of discretized differential
operators, dense factorizations) and of machine-learning inference. Analog in-memory
computing (AIMC) performs it in a single physical step: weights become the conductances of
cells in a resistive crossbar, the input vector is applied as voltages, and by Kirchhoff's
laws the bit-line currents sum to the matrix--vector product. For MVM-bound workloads the
efficiency gain is large enough to make AIMC a seriously pursued accelerator substrate.

Unlike a digital low-precision format, whose error is deterministic rounding a numerical
analyst can bound, an AIMC crossbar realizes arithmetic with a \emph{stochastic} error budget:
the programmed state is perturbed by device noise, and the analog summation adds
IR-drop and analog-to-digital conversion error on top. The designer's knob is therefore not
the rounding mode but the \emph{encoding}: how to represent each value across the physical
cells so that this hardware noise does the least damage to the result. Choosing that encoding
for maximum accuracy on an imprecise unit is exactly the question this workshop poses of
low-precision hardware in general; AIMC is an extreme, and increasingly practical, instance.

The standard encoding lever is redundancy: store each value across several cells and let the
bit-lines sum them, like the limbs of a multi-word number, except each limb is physically
noisy. This creates a genuine tension. Spreading a value across a \emph{fine} place-value grid
(a large base) buys resolution but gives the high tile a large place value, which amplifies its
noise in the summed output; storing \emph{identical} copies (base one) averages the noise away
but adds no resolution. The
base of the representation trades these off. The closest prior analysis, Le~Gallo
et~al.~\cite{legallo2022}, derives a closed-form slicing error for analog crossbars but models
device \emph{noise only}; that term is monotone in the base, so their closed form gives an
optimal base of one (equal-fill), and the resolution axis never enters. Their experiments do
report a case where $b{=}2$ helps, but through an \emph{error-correction} programming algorithm,
not from the representation itself, and without a model of why. Others fix the base
empirically~\cite{diware2021} or search it per device; Mackin et~al.~\cite{mackin2022}
optimize a significance factor by evolutionary search over a proxy metric, reporting a value of
two. What is missing is the resolution--noise \emph{trade}: an error model that also carries the
finite-grid quantization floor, so that a nontrivial base is optimal for the encoding itself and
can be read off from the device noise rather than fixed or searched. We build that model on the
assumption that each tile's MVM-output noise is \emph{additive}: its magnitude does not scale
with the stored code, though it is calibrated per tile from the programmed weights; the
non-additive parts of a real crossbar (IR-drop, ADC) are not modeled but are carried in
the device-physics simulation of Sec.~\ref{sec:eval}, where the predicted optimum is shown to
survive them.

We resolve this by treating the crossbar as a custom fixed-point representation with a
per-tile output-noise parameter~$\eta$ and asking the classical numerical question: how to represent
for minimum error under a hardware error model. We contribute:
\begin{itemize}
\item \textbf{An error model with the resolution--noise trade.} We extend the closed-form
slicing-error model of~\cite{legallo2022}, whose noise term we recover exactly, by adding the
finite-grid quantization floor, and minimize it to a closed-form optimal base,
$\bopt = 1 + 1/(12\,\bound^2\,\eta^2)$. It is an \emph{interior} optimum that recovers the prior
equal-fill ($b{=}1$) result only in the noise-dominated limit (Sec.~\ref{sec:theory}).
\item \textbf{An encoding and a search-free calibration rule.} An \emph{overlapping} place-value
split realizes the interior base while keeping the high tile's place value, and thus the
amplification of its noise in the output, small; a fixed shared rotation makes the Gaussian model
apply to real operators
(Sec.~\ref{sec:method}). The base follows from one device-noise measurement, replacing the
per-model numerical search prior work needs, at no runtime or programming cost over equal-fill.
\item \textbf{Validation across three fidelities.} From an analytic model through a full
device-physics simulator to end applications, the predicted base gives $+3.3$~dB over the naive
nibble split and is operator-agnostic across scientific and neural-network MVMs
(Sec.~\ref{sec:eval}).
\end{itemize}

\section{Low-Precision MVM as a Numerical Problem}\label{sec:background}
\paragraph*{Tile MVM and encoding} A crossbar tile computes $y = Wx$ with
$W\in\mathbb{R}^{T\times T}$. Each weight is not stored as a real number but encoded as $n_W$
integer cell states $\{s_k\}_{k=0}^{n_W-1}$, $s_k\in[-\bound,\bound]$ with
$\bound = 2^{q-1}-1$ for a signed $q$-bit cell, and decoded by a place-value sum with a shared
per-column scale,
\begin{equation}
\hat w \;=\; \mathrm{scale}\cdot\sum_{k=0}^{n_W-1} b^k\, s_k .
\label{eq:decode}
\end{equation}
The base $b$ and cell count $n_W$ define the representation: $b=1$ is equal-fill duplication,
$b=2^{q}$ is a non-overlapping bit-slice, and intermediate $b$ is the overlapping regime this
paper studies.

\paragraph*{The error budget} Two sources corrupt $\hat w$, mirroring the on-silicon
decomposition of the MVM error into a weight-programming (``linear'') term and a residual
device term~\cite{legallo2023}. \emph{Quantization} is deterministic, from the finite integer
grid, familiar from any fixed- or floating-point format (posits, custom low-precision) and
here set by the master step $\Delta=\mathrm{scale}$.
The distinctively analog source is \emph{read noise}. The output is read as $n_W$ separate tile
products $\tilde y_k$ that the periphery scales by place value $b^k$ and sums,
$\hat y = \mathrm{scale}\sum_k b^k \tilde y_k$; each tile read carries an additive error
$\tilde y_k = y_k + n_k$, with $n_k\sim\mathcal{N}(0,\sigma_{\mathrm{out}}^2)$. We normalize this
as $\eta$, the per-tile output-noise std relative to the per-tile signal RMS, the same kind of
normalized noise ($\sigma_{\delta G}/G_{\max}$) used to characterize PCM
crossbars~\cite{joshi2020}. Because the
place-value scaling is applied \emph{after} the noisy read, the output error decomposes as
\begin{equation}
\hat y - y \;=\; \underbrace{(\hat y_{\mathrm q} - y)}_{\text{quantization}}
\;+\; \underbrace{\mathrm{scale}\textstyle\sum_k b^k n_k}_{\text{noise}},
\label{eq:errdecomp}
\end{equation}
where $\hat y_{\mathrm q}$ is the noiseless quantized output. The tile at place value $b^k$ thus
has its output noise weighted by $b^k$ in the sum. This coupling of place value to output noise is
the mechanism the encoding must manage, and the object we optimize in Sec.~\ref{sec:theory}. We
measure the resulting error as the relative MVM MSE,
$\relMSE = \mathbb{E}\|y-\hat y\|^2 / \mathbb{E}\|y\|^2$, equivalently
SNR $= -10\log_{10}\relMSE$.

\paragraph*{Assumptions} For analytic tractability, and following common practice in AIMC
noise modeling, we take $W_{ij}, x_j$ i.i.d.\ Gaussian, clip weights at
$\Wmax = k_\sigma\,\sigma_W$ ($k_\sigma=3$, so clipping loss is negligible;
Sec.~\ref{sec:theory}), and use signed 4-bit cells ($\bound=7$) throughout. Crucially, we take each
tile's MVM-output noise $n_k$ to be \emph{additive}: its variance $\sigma_{\mathrm{out}}^2$ is a
per-tile quantity, calibrated from the programmed weights, but within a tile it does not scale with
the individual stored code. This models the regime in which the dominant read error is a fixed
jitter set by the ADC and peripheral circuitry rather than by the individual programmed
conductance: in a differential pair the two devices are read against each other so a common
conductance-scale factor cancels, and after global drift compensation the residual is well
approximated as additive~\cite{legallo2023}. Real crossbars are not exactly additive: the output
noise can scale with the stored code, and a conductance-proportional (multiplicative) component
would recover the
$b{=}1$ optimum of~\cite{legallo2022}. Our result applies to the additive-limited regime, with the
general case a mixture (Sec.~\ref{sec:conclusion}). As Sec.~\ref{sec:theory}
shows, it is precisely the additive term that moves the optimum off $b{=}1$. The Gaussian operand assumption is likewise not merely convenient: a fixed random
orthogonal rotation of $W$ (Sec.~\ref{sec:method}), which cancels in the MVM, drives the
per-column operand distribution toward Gaussian even for heavier-tailed real weights, so the
analysis applies to real operators, not only synthetic ones.

\section{Overlapping Base-$N$ Representation}\label{sec:method}
\paragraph*{Encoding} For $n_W=2$, a weight $w$ is master-quantized onto an integer grid of
range $\bound\,(1+b)$ and split by place value across two cells:
\begin{equation}
m = \mathrm{round}(w/\mathrm{scale}),\quad
s_H = \operatorname{trunc}(m/b),\quad
s_L = m - b\,s_H,
\end{equation}
with $\mathrm{scale} = \Wmax/[\bound\,(1+b)]$ and both $s_H,s_L$ clipped to $[-\bound,\bound]$;
the bit-lines then reconstruct $\hat w$ via the decode sum~\eqref{eq:decode}. The base $b$ is a
positive \emph{integer}: it is the radix that carves the integer code $m$ into the two integer
cell codes, so $s_L=m-b\,s_H$ is integral only for integer $b$. Absent clipping
the low digit satisfies $|s_L|\le b-1$, so for $b\le\bound+1$ it never saturates.
The departure from equal-fill is that $s_H$ and $s_L$ are in general \emph{different} integers.

\begin{figure}[t]
  \centering
  \includegraphics[width=\columnwidth]{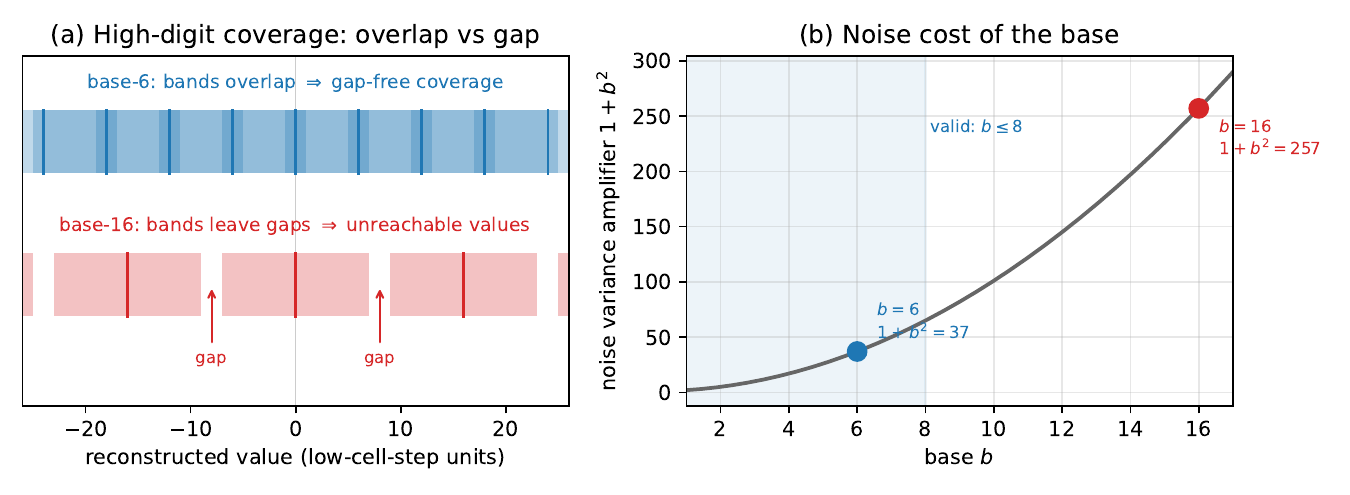}
  \caption{Overlap principle. (a)~Value-range covered by each high digit ($[k b-\bound,\,k b+\bound]$
  in low-cell-step units): for base-6 the bands overlap into gap-free coverage, while base-16's
  bands leave gaps (e.g.\ $\pm8$) in the representable grid.
  (b)~Noise cost: the tile noise variance amplifier $1+b^2$ (Eq.~\eqref{eq:noisevar}) is $37$ for
  base-6 vs.\ $257$ for base-16, the $\approx7\times$ gap behind the measured $+3.3$~dB, shaded
  over the validity range $b\le\bound{+}1$.}
  \label{fig:overlap}
\end{figure}

\paragraph*{The overlap principle} High digit $k$ covers the integers $[k b-\bound,\,k b+\bound]$,
a run of $2\bound{+}1$ values spaced $b$ apart, so the reconstructed grid has no gap precisely
when $b\le 2\bound{+}1$. Base-16 exceeds this ($16>15$): the value $8$ needs $s_L{=}8$ or $-8$,
both outside $[-7,7]$, so it is unreachable, a gap in the representable grid.
The stronger condition $b\le\bound{+}1$ used below also keeps the low digit unsaturated
($|s_L|\le b-1\le\bound$). Overlap is redundancy: the master range
$\bound\,(1+b)$ is covered \emph{without} the high tile's place value $b$ growing large. The
per-tile output noise is the same at any base; the base only sets how strongly the high tile's
noise is weighted in the summed output~\eqref{eq:errdecomp}, so the noise variance scales as
$\sum_k b^{2k}$: $1{+}6^2{=}37$ for base-6 versus $1{+}16^2{=}257$ for the naive non-overlapping
base-16 nibble split (Fig.~\ref{fig:overlap}), a $7\times$ reduction that
Sec.~\ref{sec:theory} makes exact. The base therefore sets a resolution--noise trade:
larger $b$ gives a finer grid but more strongly amplifies the high tile's noise in the output.

\paragraph*{Preconditioning} The rotation of Sec.~\ref{sec:background} is a fixed
block-diagonal random orthogonal matrix, shared across tiles and reproducible from a seed, so
it adds no storage and costs one matmul per tile block; applied to $W$ and undone on the
input, it cancels in the exact product. Beyond Gaussianizing the per-column marginal it
decorrelates the operand across the tile, the analog of incoherence preconditioning in
randomized numerical linear algebra.

\section{Closed-Form Error Analysis}\label{sec:theory}

\subsection{Two-cell noise propagation}
We take $n_W=2$ throughout this subsection (the general case is Sec.~\ref{sec:theory}\,C). The high
and low cells are read as two separate tile products that the periphery combines by place value;
we model each tile's output as carrying an \emph{additive} error $n_k$ once its code is programmed,
\begin{equation}
n_k\sim\mathcal{N}(0,\sigma_{\mathrm{out}}^2),\qquad k\in\{H,L\},
\label{eq:noisystate}
\end{equation}
where $\sigma_{\mathrm{out}}^2$ is calibrated per tile from the programmed weights but, being
\emph{additive}, does not scale with the individual stored code. On a real crossbar this output
noise can scale with the code; we assume it additive and validate the resulting prediction against
the full non-additive physics in Sec.~\ref{sec:eval}. For the closed form we take a common
$\sigma_{\mathrm{out}}$ across the high and low tiles, an idealization that the device-physics
sweep, which uses the true per-tile noise, relaxes. Combination scales the high tile by its place
value $b$, so the noise on the reconstructed output is $\mathrm{scale}\,(b\,n_H + n_L)$, with
variance
\begin{equation}
\mathrm{Var}(\hat y)_{\text{noise}} = \mathrm{scale}^2\,\sigma_{\mathrm{out}}^2\,(1+b^2).
\label{eq:noisevar}
\end{equation}
The place value enters \emph{before} this sum, which is what makes the noise grow with $b$; had a
single additive error been injected on the combined output, it would carry no $b$-dependence and
no interior optimum would exist. The variance is \emph{independent of the stored integers}
$s_H,s_L$: the integers set the noiseless signal but
never multiply the noise. This is the key observation: the MVM noise depends only on the
cells' place values $\{1,b\}$, not on whether the cells carry identical (equal-fill) or
different (place-value) values, so non-equal-fill and equal-fill share the same noise
expression. The relative MSE is this variance over the signal power. Substituting
$\mathrm{scale}=\Wmax/[\bound(1+b)]$ and $\Wmax=k_\sigma\sigma_W$, the $b$-dependence is
$k_\sigma^2\,(1+b^2)/(1+b)^2$ times a factor set by $\sigma_{\mathrm{out}}$ and the signal scale;
collecting that factor into the normalized noise $\eta$ (the per-tile output-noise std relative to
the per-tile signal RMS) gives the relative noise MSE $k_\sigma^2\eta^2\,(1+b^2)/(1+b)^2$. Only
the $b$-dependence matters for the optimum, and $\eta$ is the single device number that sets it.
This noise term coincides with
the equal-fill slicing-error expression of Le~Gallo et~al.~\cite{legallo2022} (their Eq.~8,
generalized to $n_W$ cells in Eq.~\eqref{eq:relmseN}); our contribution is not this term but the
\emph{quantization floor} we add to it next, which turns their monotone $b{=}1$ optimum into an
interior one.

\subsection{Total error and the optimal base}
The finite master grid, of step $\Delta=\mathrm{scale}$, adds a uniform-quantization term
$\Delta^2/12$ which normalizes to $k_\sigma^2/[12\,\bound^2(1+b)^2]$ (the Gaussian-tail
clipping term, the mass beyond $k_\sigma\sigma_W$, is ${\approx}4\times10^{-4}$ at
$k_\sigma=3$ and is dropped).
The total relative MVM error is
\begin{equation}
\relMSE(b) = k_\sigma^2\eta^2\frac{1+b^2}{(1+b)^2}
           + \frac{k_\sigma^2}{12\,\bound^2\,(1+b)^2}.
\label{eq:relmse2}
\end{equation}
Setting $\mathrm{d}\,\relMSE/\mathrm{d}b = 0$ yields the closed-form optimum
\begin{equation}
\boxed{\;\bopt = 1 + \dfrac{1}{12\,\bound^2\,\eta^2}\;}
\qquad(\text{capped at } 2\bound+1).
\label{eq:bopt}
\end{equation}
Since $b$ is an integer (Sec.~\ref{sec:method}), the deployed base is the nearest integer to this
continuous optimum. The minimum is shallow, so this rounding costs little: for a representative
$\eta{=}0.02$ the formula gives $\bopt{=}5.25$, and both integer neighbors, $b{=}5$ and $6$, lie
within $0.02$~dB of the continuous optimum, with the whole band $b{=}3$--$8$ within $0.2$~dB. The
continuous formula thus selects which integer bases to consider; the flat minimum makes the final
rounding immaterial, so the calibration need only land in the basin, not hit an exact value.
Two limits interpret~\eqref{eq:bopt}, and one of them recovers the prior result. As $\eta\to0$
the optimum runs to the cap (precision-limited: use the finest master grid, i.e.\ pure
bit-slicing). As $\eta\to\infty$ it falls to $b=1$, equal-fill duplication, whose averaging is
optimal when noise dominates. This $b{=}1$ limit is exactly the conclusion of
Le~Gallo et~al.~\cite{legallo2022}, who minimize the noise term alone (the first summand
of~\eqref{eq:relmse2}, monotone in $b$): their PCM noise is state-dependent (multiplicative), so
the resolution axis never enters their closed form and ADC quantization is handled separately
from the base. (Their experiments do find $b{=}2$ optimal in one case, but via an
error-correction programming algorithm rather than the representation itself.)
Folding the master-grid quantization floor into the same objective is what creates the interior
optimum at moderate $\eta$: under \emph{additive} output noise the resolution gained by a
finer grid is worth the extra amplification of the high tile's noise up to $\bopt$. The two conclusions
are thus the two regimes of one formula: multiplicative/noise-dominated ($b{=}1$) versus
additive/resolution-limited (an interior $\bopt$, the overlap regime of Sec.~\ref{sec:method}),
where at the optimum the scheme is noise-floor-limited with peak SNR scaling as $1/\eta^2$.

\subsection{General $n_W$ cells}
For $n_W$ cells at place values $\{1,b,\dots,b^{\,n_W-1}\}$ the same derivation gives
\begin{equation}
\relMSE = k_\sigma^2\eta^2\,\frac{T}{S^2} + \frac{k_\sigma^2}{12\,\bound^2\,S^2},
\label{eq:relmseN}
\end{equation}
with the geometric sums $S=(b^{n_W}-1)/(b-1)$ and $T=(b^{2n_W}-1)/(b^2-1)$. The noise factor
$T/S^2$ is algebraically identical to the equal-fill slicing error of Le~Gallo
et~al.~\cite{legallo2022} (their Eq.~8, squared); we recover it and add the quantization floor.
The corners of~\eqref{eq:relmseN} recover known results: $b=1\Rightarrow T/S^2=1/n_W$
(equal-fill: error drops by
exactly $n_W$, the familiar noise-averaging gain); $n_W=1\Rightarrow T/S^2=1$ (single-cell baseline);
$b\to\infty\Rightarrow T/S^2\to1$ (noise-averaging benefit vanishes; pure precision extension).

\emph{Does the optimum change structurally with $n_W$?} The tradeoff itself does not: the noise
factor $T/S^2$ still rises with $b$ and the quantization floor $1/(12\bound^2 S^2)$ still falls,
so~\eqref{eq:relmseN} keeps an interior minimum. What changes is where it sits. Adding cells
supplies resolution through \emph{more slices} rather than a finer per-step grid: $S$ grows as
$b^{\,n_W-1}$, so the quantization term collapses far faster in $n_W$ than in $b$, and its
gradient in $b$ flattens (numerically, $\mathrm{d}(1/S^2)/\mathrm{d}b$ at fixed $b$ shrinks
sharply from $n_W{=}2$ to $3,4$). The noise term, by contrast, keeps rising in $b$. The balance
therefore tips toward \emph{smaller} bases as $n_W$ increases: at $\eta{=}0.02$ the best integer
base in $[1,2\bound{+}1]$ moves from $b{=}5$ ($n_W{=}2$) to $b{=}2$ ($n_W{=}3,4$). So more cells
do not call for a larger base; they let a smaller, less noise-amplifying base reach the same
resolution. The $n_W{=}2$ case we deploy is the one where the base is doing the most work.

\subsection{Where non-equal-fill wins}
Since the two encodings share the identical noise term (Eq.~\eqref{eq:noisevar}), the comparison
between equal-fill and place-value reduces entirely to the quantization floor: for the same cell
budget, place-value provides $S=1+b$ reconstruction levels versus $n_W$ for equal-fill.
Place-value therefore strictly wins whenever quantization is non-negligible, the few-bit regime
AIMC occupies, which is the same statement, from the encoding side, as $\bopt>1$: both say the
resolution term matters. The two coincide only at $b=1$, when noise dominates and quantization is
irrelevant.

\section{Evaluation}\label{sec:eval}
The theory makes a falsifiable prediction: when the tile-output noise is \emph{additive}
(does not scale with the stored code), the optimal base is \emph{not} one, opposing the equal-fill ($b{=}1$) optimum
that holds under multiplicative noise~\cite{legallo2022}. We therefore test not just whether the
closed form fits, but which regime the hardware occupies, and find the interior optimum, at three
levels of fidelity.

\begin{figure}[t]
  \centering
  \includegraphics[width=\columnwidth]{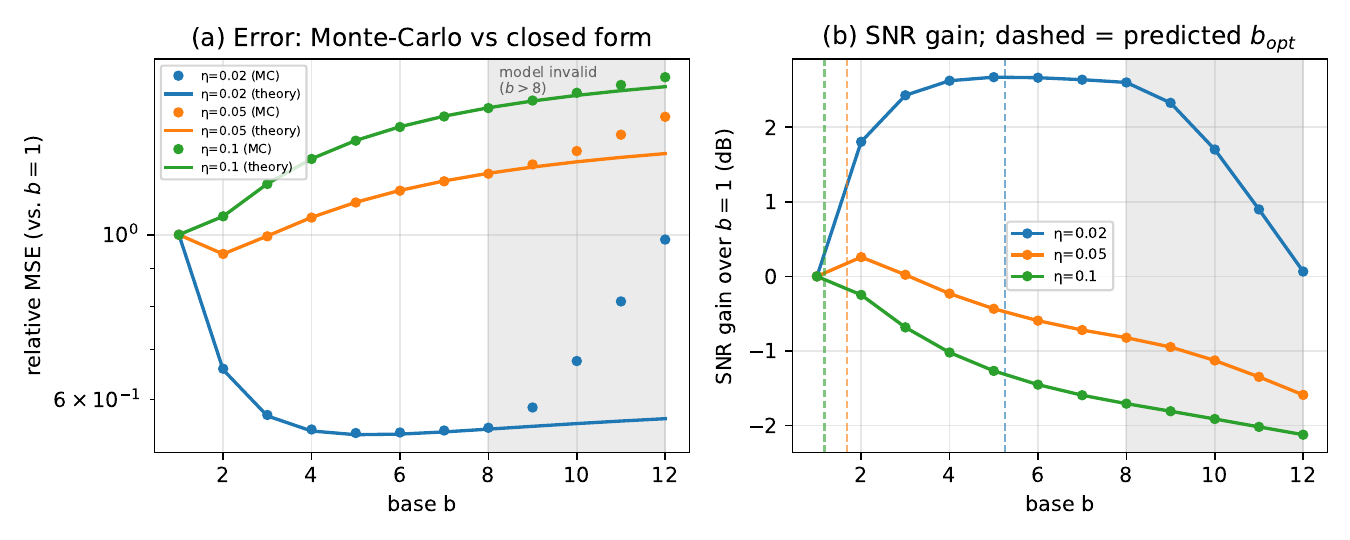}
  \caption{Theory validation. (a)~relative MSE vs.\ base, Monte-Carlo (markers) vs.\ closed
  form~\eqref{eq:relmse2} (solid). (b)~SNR; dashed verticals mark the
  predicted $\bopt$ from~\eqref{eq:bopt}. Shading marks $b>\bound{+}1$, beyond the analytic
  validity bound, where Monte-Carlo is expected to depart from theory.}
  \label{fig:theory}
\end{figure}

\paragraph*{Analytic (Monte-Carlo)} Fig.~\ref{fig:theory} sweeps the base for
$\eta\in\{0.02,0.05,0.10\}$: Monte-Carlo relative MSE (markers) tracks the closed
form~\eqref{eq:relmse2} (curves), and the SNR peaks fall on the predicted
$\bopt=\{5.25,1.68,1.17\}$. Agreement is within ${\approx}1\%$ for $b\le\bound+1$; beyond that
(shaded) the low cell cannot hold every residue and the model's validity bound is exceeded,
seen as the Monte-Carlo curve rising above theory, a departure confined to the region the
analysis already excludes.

\begin{figure}[t]
  \centering
  \includegraphics[width=0.86\columnwidth]{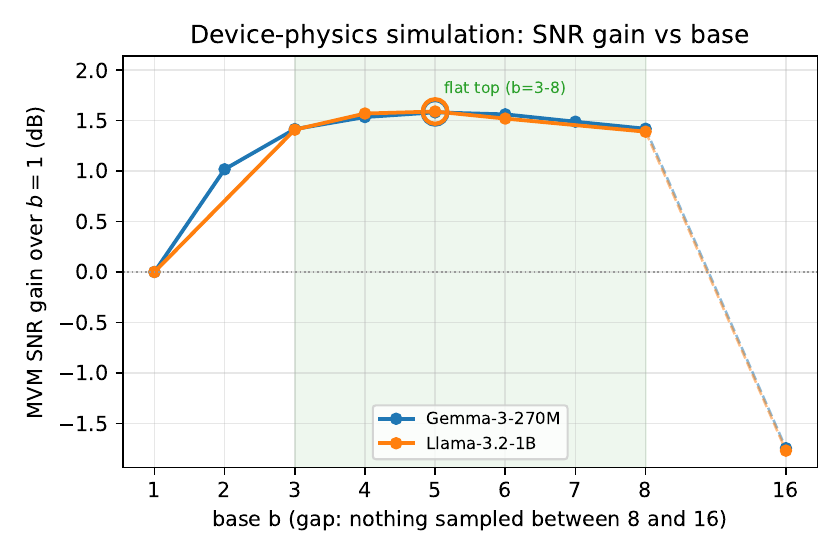}
  \caption{Device-physics simulation: mean per-layer MVM SNR vs.\ base (IR-drop, programming
  noise, ADC) for Gemma-3 270M and Llama-3.2 1B. Both trace the same inverted-U, peaking at
  $b{=}5$ (circled) with a flat top ($b{=}3$--$8$) and ${\approx}3.3$--$3.4$~dB over base-16 (a
  separated point, dashed; nothing sampled in $8$--$16$).}
  \label{fig:fullsim}
\end{figure}

\paragraph*{Device-physics simulation} We evaluate on a physics-based simulator of an analog
in-memory crossbar tile. As one representative point in the AIMC design space (among many
possible array sizes and cell precisions) we model a $512{\times}512$ array of multi-level cells
storing signed 4-bit weights differentially; the method and its closed form are not specific to
this choice. The simulator reproduces the dominant non-idealities of AIMC MVM:
programming variation, IR-drop from finite wire and cell resistance solved on the
resistive network, and ADC quantization of the bit-line currents. The resulting tile-output noise
is \emph{not} the additive quantity our closed form assumes: IR-drop is a
nonlinear function of the currents actually flowing and ADC error depends on the output
magnitude, so the effective output noise scales with the code. This makes the test stronger than the
model, which asks whether the predicted optimum survives once realistic, non-additive output
effects replace the idealized additive assumption. Fig.~\ref{fig:fullsim} reports mean per-layer
MVM SNR \emph{gain relative to} $b{=}1$ under this model (absolute levels are omitted). It does
survive: the SNR does \emph{not} peak at $b{=}1$ but rises to an interior maximum at $b{=}5$,
${\approx}1.6$~dB above $b{=}1$, and beats the naive base-16 split by ${\approx}3.3$--$3.4$~dB.
Both models trace the same inverted-U and peak at the same base, so the optimum is model-agnostic.
The interior optimum predicted from additive output noise thus holds under the full IR-drop and
ADC physics, so on this device the base should be tuned, not fixed at one. The effective $\eta$
is not a single datasheet constant: it aggregates programming noise, IR-drop, ADC quantization,
per-column normalization, and the operand distribution, so it must be \emph{calibrated} for a
given device and workload rather than derived in closed form. The device-physics simulation is
that calibration: the peak location fixes the operating $\eta$ ($b{=}5$ inverts
through~\eqref{eq:bopt} to $\eta\approx0.02$), and this is the ``single device-noise measurement''
the encoding needs. What the closed form contributes is the \emph{shape} that measurement is read
against, an interior optimum with a broad flat top ($b{=}3$--$8$) rather than the monotone descent
to $b{=}1$ of the multiplicative regime; the exact base within that basin is not critical, so the
calibration need only land in it.

\emph{Why a single global base.} Since $\eta$ is a per-tile quantity, $\bopt$ is in principle
per-tile; a global base is justified only if the tiles agree. They do. Sweeping the base
\emph{per} $512{\times}512$ tile and recording each tile's best base, the optimum is tightly
concentrated: on Gemma $48\%$ of tiles peak at $b{=}5$ and $92\%$ within $b{\in}\{4,5,6\}$; on
Llama $46\%$ at $b{=}5$ and $97\%$ within $\{4,5,6\}$; essentially no tile prefers a base outside
that band. No tile statistic predicts the optimum (correlations of the per-tile best base with
weight std, kurtosis, and dynamic range are all $|r|{<}0.05$), and the per-tile base-sensitivity
is only $0.22$~dB. The rotation (Sec.~\ref{sec:method}) is what homogenizes the tiles: it drives
every tile's operand toward the same Gaussian, so all tiles want the same base and a per-tile or
learned base rule has nothing to exploit. A single global $b{=}5$--$6$ is therefore near-optimal
tile-by-tile, not merely on average.

\paragraph*{End-application fidelity} The optimum is a property of the encoding under noise, so
it is \emph{operator-agnostic}. On a single matrix--vector product for representative
\emph{scientific} operators, a 2D five-point Laplacian (finite-difference/FEM stencil) and an
ill-conditioned dense matrix ($\kappa{=}10^3$), the same inverted-U appears, peaking at
$b{\approx}4$--$6$ with ${\approx}3.4$~dB over base-16 (Table~\ref{tab:apps}). On transformer inference
we compare three encodings at the same two-cell budget: \emph{equal-fill} ($b{=}1$) stores the
same value in both cells and averages; \emph{rotated non-uniform} applies the fixed rotation of
Sec.~\ref{sec:method} and quantizes both cells to shared Lloyd-Max levels, an incoherence-style
baseline in the spirit of~\cite{tseng2024,liu2024}; \emph{residual redundancy} puts the weight in
one cell and its leftover residual in the other, the successive error-correction strategy with
which~\cite{legallo2022} reaches an interior base. Base-$N$ split (at $b{=}6$, the integer neighbor
of the $\bopt{\approx}5$ of Sec.~\ref{sec:theory}) clearly beats equal-fill and rotated non-uniform,
and \emph{matches} residual redundancy, the strongest baseline, to within the run-to-run noise on
both models ($\pm0.008$). The distinction is not accuracy but cost: base-$N$ reaches the
top group using only \emph{uniform} placement, whereas residual redundancy needs per-tile
Lloyd-Max codebooks, and unlike either baseline the base follows from the closed-form optimum
rather than empirical search. Accuracy is averaged
over eight zero-shot commonsense tasks from the LM-Evaluation-Harness (ARC-easy, ARC-challenge,
HellaSwag, PIQA, WinoGrande, OpenBookQA, BoolQ, COPA), each a ten-run mean with noise resampled
per run. Notably
equal-fill ($b{=}1$), the optimum for multiplicative noise, is the
\emph{worst} row on both models, the end-application signature of the additive regime the
theory identifies. The flat basin also shows up here: re-running base-$N$ at $b{=}5$ (the other
integer neighbor of $\bopt{=}5.25$, with its own matched calibration) changes the eight-task
accuracy by $+0.0003$ on Gemma (indistinguishable from $b{=}6$) and $-0.006$ on Llama (within
about one standard deviation, $\pm0.005$). Either base is near-optimal in accuracy, so the
rounding of the continuous $\bopt$ to an integer is immaterial in practice.

\begin{table}[t]
  \centering
  \caption{End-application fidelity, reported \emph{relative to equal-fill} ($b{=}1$) so no
  absolute capability is disclosed: Laplacian single-matvec SNR \emph{gain} over $b{=}1$ (dB), and
  transformer eight-task accuracy as \emph{\% change} over $b{=}1$ (10-run means, $\pm0.008$
  absolute run-to-run noise on the accuracy). All encodings use two cells per weight; rotated
  non-uniform and residual redundancy are data-dependent neural-network encodings with no fixed
  scientific-matvec analog, hence the omitted Laplacian entries. Base-$N$ split ($b{=}6$) matches
  the strongest baseline (residual redundancy) within noise, using only uniform placement.}
  \label{tab:apps}
  \begin{tabular}{lccc}
    \toprule
    & Laplacian & \multicolumn{2}{c}{8-task acc.} \\
    & SNR gain & \multicolumn{2}{c}{($\Delta\%$ vs.\ equal-fill)} \\
    \cmidrule(lr){2-2}\cmidrule(lr){3-4}
    Encoding & (dB) & Gemma-270M & Llama-1B \\
    \midrule
    equal-fill ($b{=}1$)      & $+0.0$ & $+0.0\%$ & $+0.0\%$ \\
    rotated non-uniform       & ---    & $+4.6\%$ & $+3.9\%$ \\
    residual redundancy       & ---    & $+12.3\%$ & $+7.7\%$ \\
    base-$N$ split ($b{=}6$)  & \textbf{$+1.6$} & \textbf{$+13.1\%$} & \textbf{$+9.0\%$} \\
    \bottomrule
  \end{tabular}
\end{table}

\section{Related Work}\label{sec:related}
\paragraph*{AIMC error analysis and weight slicing} Accurate inference on phase-change-memory
crossbars is well established~\cite{joshi2020}. Closest to us, Le~Gallo et~al.~\cite{legallo2022}
give a closed-form model of the analog slicing error as a function of the base and slice count,
and validate it against a PCM chip. Their closed form is \emph{noise-only}: under the
state-dependent (multiplicative) PCM noise they consider, the error is monotone in the base and
minimized at $b{=}1$ (equal-fill), and ADC quantization is treated separately. Their experiments
do surface an interior case ($b{=}2$ optimal at $t_0$ with a max-fill error-correction
algorithm), but as a programming-algorithm effect rather than a property of the representation,
and without a model of the base that would explain or locate it. We build directly on their
noise expression, which we recover exactly (Sec.~\ref{sec:theory}), and add the finite-grid
quantization floor to the \emph{same} objective. Under additive output noise this produces an
\emph{interior} optimal base in closed form, of which their $b{=}1$ closed-form result is the
noise-dominated limit; the base optimum thus depends on the device's noise regime rather than
being universally one. Complementary to the analytic view, a realized 64-core PCM accelerator
stores weights at equal significance and decomposes the measured error into programming and
residual terms~\cite{legallo2023} (our $b{=}1$ corner and error decomposition), and
Mackin et~al.~\cite{mackin2022} optimize an \emph{unequal} significance factor numerically by
evolutionary search, where our closed form removes the per-device search.
Device-variation studies quantify how programming noise degrades inference
accuracy~\cite{wang2022,bhattacharjee2023}, hardware-aware training partly recovers
it~\cite{rasch2023}, and noise-optimized rescaling adapts deployment to the device~\cite{nora2025};
advances in multi-level conductance control now offer thousands of programmable
states~\cite{rao2023}. Other bit-slicing schemes
allocate weight bits across cells: Diware et~al.\ propose an \emph{unbalanced}
bit-slicing~\cite{diware2021} and an accurate, energy-efficient RRAM bit-slicing~\cite{diware2023},
and refined multi-conductance programming improves per-cell precision~\cite{so2024}. These
choose the split empirically; we contribute the analytic optimum and its device-parameter
dependence.

\paragraph*{Low-precision numerics} Our question, how to represent values for minimum error on
an imprecise arithmetic unit, parallels the reduced-precision program in scientific computing:
mixed-precision iterative refinement~\cite{oktay2021}, alternative number
formats~\cite{klower2020}, and low-bit floating-point with error mitigation~\cite{cambier2020}.
We share the representation-optimization goal but differ in the error model: stochastic device
noise rather than deterministic rounding.

\paragraph*{Randomized preconditioning} The optional rotation is a fixed orthogonal transform
that Gaussianizes and decorrelates the operand, in the lineage of incoherence-based methods for
quantization: Hadamard-incoherence and learned-rotation schemes~\cite{tseng2024,liu2024} use the
same principle to tame outliers before low-bit encoding. A separate axis of the same encoding
problem is the choice of \emph{clip and reconstruction levels} for a single cell (uniform vs.\
non-uniform placement); here rotation and non-uniform levels enter only as the
\emph{rotated non-uniform} baseline we compare against, and our contribution is orthogonal to
them, the significance \emph{base} of the multi-cell split rather than the within-cell codebook.

\section{Conclusion}\label{sec:conclusion}
Weight splitting on a noisy in-memory accelerator is governed by a single dial. Because each
tile's additive output noise enters the sum weighted only by its place value, not by the stored
code, the base $b$ trades quantization resolution against amplified noise with no other coupling, and the
trade has a closed-form optimum $\bopt = 1 + 1/(12\,\bound^2\,\eta^2)$, an \emph{overlap} of
the cell ranges set entirely by the device noise $\eta$ and the cell resolution, with
duplication and bit-slicing as its two endpoints. The practical consequence is a reversal of
prior guidance: equal-fill ($b{=}1$) is optimal only when the tile-output noise is
multiplicative, whereas when it is additive the best base is interior, which we confirmed from an
analytic model through full device physics (with realistic, non-additive IR-drop and ADC) to
scientific and neural-network MVMs. The optimum can be determined
quickly, before device programming: measure $\eta$ once and evaluate $\bopt$ in closed form,
with no per-model numerical search. Because the base only sets how a weight is split across
cells, using $\bopt$ carries the same programming and read cost as equal-fill.

\paragraph*{Scope and limitations} The result rests on an additive tile-output-noise model
(noise that does not scale with the stored code), and its reach is exactly that model's. (i)~\emph{Mixed noise.} Real
crossbars show code-dependent output noise: a conductance-proportional (multiplicative) component
and state-dependent effects such as retention drift and programming errors that depend on the
target state; our variance $\sigma_{\mathrm{out}}^2(1+b^2)$ omits these. A multiplicative term
adds a $b$-dependent factor that pulls the optimum back toward $b{=}1$, so the true optimum for a
given device lies between our additive $\bopt$ and the multiplicative $b{=}1$; quantifying that
interpolation from a measured noise-vs-code curve is the natural extension. (ii)~\emph{Discrete base.} $\bopt$ is a
continuous relaxation of an integer parameter; we round to the nearest feasible base, which the
shallow minimum makes lossless in practice (Sec.~\ref{sec:eval}). (iii)~\emph{Validity range.}
The two-cell form holds for $b\le\bound+1$ (beyond which the low cell cannot cover every residue),
and for $n_W>2$ the optimum shifts to smaller bases (Sec.~\ref{sec:theory}\,C). (iv)~\emph{Single
matvec.} We optimize one $Wx$; error propagation through an iterative solver, where the
per-step optimum need not be the fixed-point optimum, is a separate study. More broadly, casting
an analog crossbar as a custom number representation lets the low-precision-numerics toolkit set
hardware encodings by closed form rather than by convention.

\end{document}